# Kitaev quantum spin liquid – concept and materialization

Hidenori Takagi[1,2,3*], Tomohiro Takayama[1,2*], George Jackeli[1,2*] and Giniyat Khaliullin[1*]

[1]Max Planck Institute for Solid State Research, Stuttgart 70569 Germany

[2]Institute for Functional Matter and Quantum Technologies, University of Stuttgart, Stuttgart 70569, Germany

[3]Department of Physics, The University of Tokyo, Tokyo 113-0022 Tokyo, Japan

Stephen E. Nagler*

Neutron Scattering Division, Oak Ridge National Laboratory, Oak Ridge, Tennessee 37831, USA

Abstract

A decade ago, Alexei Kitaev proposed an exactly solvable $S = 1/2$ model on a two-dimensional honeycomb lattice, where the spins fractionalize into Majorana fermions and form a topological quantum spin liquid (QSL) in the ground state. It was soon recognized that a family of complex iridium oxides, as well as ruthenium chloride, with honeycomb structure are magnetic insulators and accommodate essential ingredients of the Kitaev model, due to the interplay of electron correlation and spin-orbit coupling. This initiated a race to materialize the Kitaev QSL and to capture the signature of Majorana fermions. In this review, we provide a wide perspective of this rapidly growing field, including theory, materials and experiment. We first summarize the theoretical background of the Kitaev QSL ground state and its materialization using spin-orbital-entangled $J_{\text{eff}} = 1/2$ moments. This is followed by an overview of candidate materials and their magnetic properties, including $Na_2IrO_3$, $\alpha, \beta, \gamma$-$Li_2IrO_3$, $\alpha$-$RuCl_3$ and $H_3LiIr_2O_6$. Finally, we review the latest exciting progress in the search for the Kitaev QSL. In particular, $H_3LiIr_2O_6$ and $\alpha$-$RuCl_3$ in applied magnetic field show signatures of the QSL state, and $\alpha$-$RuCl_3$ has unusual magnetic excitations and thermal transport properties that are consistent with spin fractionalization.

*e-mail: h.takagi@fkf.mpg.de; t.takayama@fkf.mpg.de; g.Jackeli@fkf.mpg.de; g.khaliullin@fkf.mpg.de; naglerse@ornl.gov

*Introduction*

In conventional magnetic materials interactions between the spins lead to a phase transition from a high temperature thermally disordered state to a magnetically ordered state at low temperature. The transition is typically accompanied by spontaneous symmetry breaking, singularities in the thermodynamic observables, and a reduction of the spin entropy to zero in a unique non-degenerate ground state. However, there exists another way to release the spin entropy without any symmetry breaking down to zero temperature, by forming a collective quantum state with long-range quantum entanglement between the spins. This exotic state of matter is called a quantum spin liquid (QSL)[1-3]. An important goal of condensed matter physics is to discover novel quantum phases formed by the ensemble of interacting spins and charges in solids. The QSL is perhaps one of the most exotic quantum phases known so far and has been attracting condensed matter scientists for a long time.

The exploration of QSLs in more than one-dimension was launched by the conjecture of the resonant valence bond (RVB) state by Philip Anderson in 1973[4]. When antiferromagnetically interacting Heisenberg $S = 1/2$ spins are placed on a triangular lattice, the interactions on different bonds conflict with each other due to the geometry, preventing spins from finding a unique way of breaking the symmetry. Anderson proposed that the ground state consists of a quantum superposition of spin singlets formed by pairs of $S = 1/2$ spins, where effectively the spins involved in the pairing fluctuate in a liquid-like fashion, in contrast to magnetically ordered ground states where the static nature of the spins conceptually resembles a solid. This gives an intuitive image of the QSL state. The ground state of the $S = 1/2$ Heisenberg antiferromagnet on the triangular lattice was later shown to have an ordered ground state, the non-collinear 120° structure[5,6]. Nevertheless, RVB-QSLs are believed to exist in other geometrically frustrated lattices such as the kagome lattice[7] or the triangular lattice with additional interactions[8-10].

Excitations in conventional magnets have $S = 1$, and generally show up in scattering experiments as peaks that are sharp in energy for a given momentum, i.e. as magnons with a well-defined dispersion. In contrast, one of the hallmarks of RVB-QSLs is the emergence of unusual elementary excitations described by mobile fermionic $S = 1/2$ quasi-particles, called spinons[1], as in the one-dimensional $S = 1/2$ Heisenberg antiferromagnet[11]. Here the conventional $S = 1$ excitation "fractionalizes" into spinon pairs. Spinons give rise to an energy continuum of excitations at a given momentum, much like electronic excitations in metals. If the spinon excitations are gapless, a Fermi surface of spinons may emerge[12,13].

A number of $S = 1/2$ triangular and kagome Heisenberg antiferromagnets have been argued experimentally to be materializations of RVB-QSL states, including organic charge-transfer salts, $\kappa$-(BEDT-TTF)$_2$Cu$_2$(CN)$_3$[14], EtMe$_3$Sb[Pd(dmit)$_2$]$_2$[15] and Herbertsmithite ZnCu$_3$(OH)$_6$Cl$_2$[16]. These compounds do not show any clear signature of magnetic ordering down to the lowest temperature measured, which is at least two orders of magnitude smaller than the energy scale of magnetic interaction. In organics, a finite density of zero-energy excitations is experimentally observed as a $T$-linear specific heat at low temperatures[17], and the associated excitations are highly mobile, as shown by a $T$-linear thermal conductivity[18]. In Herbertsmithite, the presence of continuum excitations with a small excitation gap was recently shown by inelastic neutron scattering (INS)[19] and NMR[20] measurements. These are very likely the fractionalized excitations expected for a QSL.

Despite impressive progress, research on the RVB-QSL state has been constrained by the difficulty that it has never been obtained as an exact solution of any realistic model Hamiltonian. However, recently a theoretical breakthrough was made in the field of QSLs. Alexei Kitaev proposed a simple but novel model that is exactly solvable and gives a QSL ground state, where the spins fractionalize into emergent quasiparticles - Majorana fermions[21]. Soon after, a spin-orbital $J_{eff}$ = 1/2 Mott insulator was identified in a complex iridium oxide[22]. This led to a theoretical proposal[23] for the materialization of the Kitaev model using $J_{eff}$ = 1/2 pseudo-spins in an iridate, and initiated a search for the QSL state and the hidden Majorana fermions in a family of iridium and ruthenium compounds. A new interdisciplinary field emerged, comprising of quantum magnetism, topological physics, correlated electron physics, and solid-state chemistry.

### *The Kitaev model and quantum spin liquid*

The Kitaev model consists of $S$ = 1/2 spins on a honeycomb lattice, which are coupled to the three nearest neighbors by Ising interactions with bond-dependent easy-axes parallel to the *x*-, *y*- or *z*- axes[21], as depicted in **Fig. 1a**. See **Box 1** for the corresponding Hamiltonian. The orthogonal anisotropies of the three nearest-neighbor bonds conflict with each other, giving rise to strong magnetic frustrations. In the classical limit, where quantum mechanical $S$ = 1/2 spins are replaced by a vector, like a compass needle, the ground-state manifold of the Kitaev model turns out to be infinitely degenerate[24,25]. The presence of such extensive degeneracy is a hallmark of strong frustration and is inherent to geometrically frustrated magnets such as kagome[26] and pyrochlore[27] Heisenberg antiferromagnets, the candidates for an RVB-QSL. In each configuration within the classical ground-state manifold of the Kitaev model, the honeycomb lattice is decomposed into non-overlapping nearest-neighbor 'happy' bonds with maximum exchange energy gain, achieved by aligning spins along the corresponding easy-axis. The remaining 'unhappy' bonds are highly frustrated and gain zero energy. The ground-state degeneracy is related to the choice of the distribution of the 'happy' bonds on a honeycomb lattice and among the two possible spin alignments on each of them. When quantum mechanical effects are turned on, the system starts tunneling and floating within the classical ground state manifold forming a highly entangled QSL state supporting fractionalized excitations. The true QSL ground-state can be characterized as a quantum mechanical superposition of the classical configurations, each having 1/3 'happy' and remaining 2/3 'unhappy' bonds as seen in **Fig. 1b**, in a sense somewhat similar to an RVB state.

The Kitaev model is an alternative pathway to a QSL but distinct from the others in that it is based on Ising, rather than Heisenberg interactions, and is exactly solvable[21]. By mathematically replacing the spin operator $S^\gamma$ ($\gamma = x, y, z$) with two types of Majorana operators $b^\gamma$ and $c$ as $S^\gamma = \frac{i}{2} b^\gamma c$, (**Figs. 1c and d**, and **Box 1**), Kitaev proved that the ground state is a QSL and is described as an ensemble of localized and itinerant Majorana fermions. A product of $b^\gamma$ over a bond $ij$ is defined as a bond variable operator $u_{ij}^\gamma$ and the product over a hexagon forms a $Z_2$ gauge flux (sometimes called "vison") operator $W$. $u_{ij}^\gamma$ and $W$, with eigenvalues $\pm i$ and $\pm 1$ respectively, commute with the Hamiltonian and therefore are conserved. The Majorana fermions associated with $b^\gamma$ are immobile due to the conservation but control the sign of hopping of the Majorana fermions associated with $c$, as shown in **Fig. 1d**. In the ground state, the signs of all bond variables and fluxes are equal (**Fig. 1e**), giving rise to a coherent motion of $c$ Majorana fermions with a Dirac dispersion, depicted in **Fig. 1f,** analogous to that of an electron in graphene[28]. The emergence of the itinerant Majorana fermions is nothing but a fractionalization of $S$ = 1/2 spins. See **Box 1** for details. The ground state forms a $Z_2$ QSL with gapless fermionic excitations, and due to the absence of a gap, is considered quasi-topological.

The elementary excitations of the Kitaev QSL should mirror the fractionalization of spins. Following the original exact solution, various physical observables such as the dynamic spin structure factor[29] and Raman response[30,31] have been computed exactly. The low-energy spin excitations are localized as they comprise of not only itinerant but also immobile Majorana fermions, which emerge as a **Q**-independent spin response with an excitation gap in the spin structure factor[29] (the spin response for a perturbed Kitaev model is discussed in the last section of this review). The low-energy Raman response captures the fermionic excitations as the result of fractionalization. Thanks to a newly-implemented numerical algorithm, specially designed for Majorana systems, the thermodynamics of the model became accessible over almost the full temperature range[32]. In the specific heat $C(T)$, the fractionalization of spins shows up as the presence of two well-separated peaks: one at a high temperature corresponding to the itinerant Majorana fermions and the other at a low temperature corresponding to flux ordering of the localized Majorana fermions. Each peak carries an entropy of 50% of $R$ln2. A half-quantized thermal Hall effect $\kappa_{xy}$ associated with the chiral edge state of Majorana fermions, $\kappa_{xy}/T = z(\pi/6)(k_B^2/\hbar)$, was theoretically predicted[21,33], where $z$ is a half integer number and $k_B$ and $\hbar$ are the Boltzmann and Planck constants.

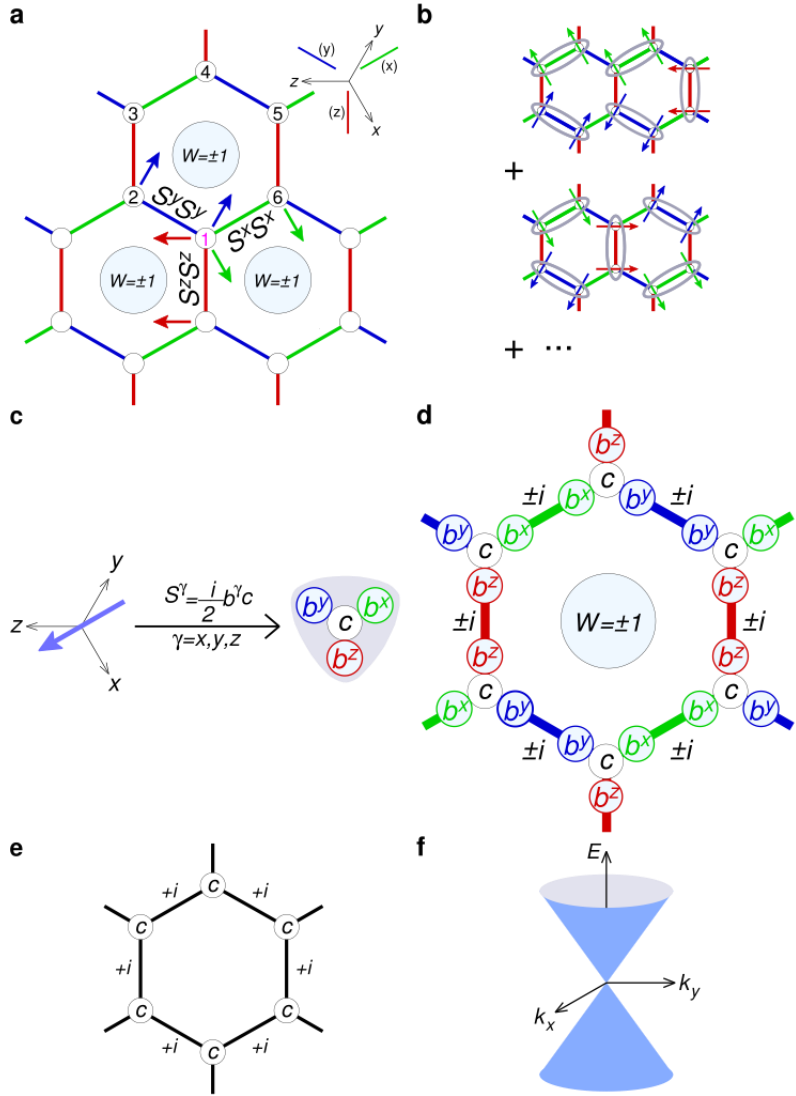

**Figure 1. The Kitaev model. a,** The $S = 1/2$ spins on a honeycomb lattice with bond dependent Ising interactions. The green, blue and red bonds have the Ising easy-axes parallel to the $x$-, $y$- or $z$- axes respectively. Product of six spins around each hexagonal loop forms a conserved quantity, e.g. $W_{1-6} = 2^6 S_1^z S_2^x S_3^y S_4^z S_5^x S_6^y$ with eigenvalues $W = \pm 1$, defining $Z_2$ flux through the hexagons. **b,** An image of Kitaev quantum-spin-liquid state as a quantum superposition of entangled states with different distribution of spin paired `happy' bonds. **c,** A real spin fractionalizes into two kinds of Majorana fermions. "$b^\gamma$" represents the three localized Majorana fermions and "$c$" represents the one itinerant Majorana fermion. **d,** Emergent Majorana fermions on the honeycomb lattice. The solid lines depict the conserved bond variables $u_{ij}^\gamma = b_i^\gamma b_j^\gamma$ of localized fermions with their imaginary eigenvalues $\pm i$ which controls the phase of $c$ fermion hopping amplitude. Their product around each hexagonal loop measures the $Z_2$ flux as $W_{1-6} = u_{12}^y u_{23}^z u_{34}^x u_{45}^y u_{56}^z u_{67}^x$. **e,** In the ground state, the hopping amplitudes have uniform phase, e.g. $+i$, allowing a coherent propagation of $c$ fermion to gain a Dirac dispersion shown in **f.**

## Materialization of the Kitaev model

The Kitaev model was originally thought to be a toy model for theorists because pure $S = 1/2$ spins in general do not accommodate strong Ising anisotropy. However, heavy $4d$ and $5d$ transition metal compounds have recently emerged as a new paradigm for correlated electron physics that may realize Kitaev behavior due to an interplay between correlation and strong spin-orbit coupling[34]. Among them, $Ir^{4+}$ oxides and a $Ru^{3+}$ chloride with $d^5$ electronic configuration and rock-salt-related honeycomb structures turned out to be a promising playground for the Kitaev QSL.

In $Ir^{4+}$ oxides with octahedral coordination of oxygen ions, a large crystal-field splitting of 3 - 4 eV produced by the oxygen octahedron wins over the Hund's first rule which favors the high spin state. All the five $d$ electrons are therefore accommodated in the triply degenerate $t_{2g}$ manifold comprising of $d_{xy}$, $d_{yz}$, and $d_{zx}$ orbitals as seen in **Fig. 2a**. Spin-orbit coupling $\lambda_{SO}$, relativistic in origin, becomes as large as a half eV for heavy elements like Ir and is much larger than the typical crystal-field splitting of the $t_{2g}$ manifold of 0.1-0.2 eV, produced by cubic-symmetry breaking distortions of oxygen octahedra. The effect of spin-orbit coupling can be introduced by treating the $t_{2g}$ manifold as if it were a triply degenerate manifold of $p$-orbitals, with orbital angular momentum $l = 1$. The $l = 2$ angular-momentum operator of atomic $d$ levels, projected into the $t_{2g}$ manifold, acts as an effective angular momentum $l_{eff} = 1$ but with a minus sign, $l = -l_{eff} = -1$[35]. Taking only the dominant spin-orbit coupling into consideration, the $t_{2g}$ manifold splits into the higher $j_{eff} = 1/2$ Kramers doublet and the lower $j_{eff} = 3/2$ quartet with a splitting energy of $3\lambda_{SO}/2$. With five $t_{2g}$-electrons, the $j_{eff} = 3/2$ quartet is fully occupied while the $j_{eff} = 1/2$ Kramers doublet accommodates one electron or equivalently one hole. As there is only one hole in the $t_{2g}$ manifold, it is often convenient to describe the electronic configuration in a "hole" picture as shown in **Fig. 2b**. Upon introducing hopping between the neighboring sites, a half-filled and relatively narrow band derived from the $j_{eff} = 1/2$ doublet is formed. A modest on-site Coulomb repulsion $U$ of ~2 eV can open a charge gap in the half-filled band produced by the strong spin-orbit coupling[36] and as a result $Ir^{4+}$ oxides are often weak spin-orbital Mott insulators with $j_{eff} = 1/2$ moments. Such a state was first identified in the layered perovskite $Sr_2IrO_4$[22].

Starting from the limit of strong Hund's coupling (the *LS*-coupling scheme) instead of strong spin-orbit coupling (the *jj*-coupling scheme described above), the six-fold degenerate manifold of five localized $t_{2g}$ electrons, with total spin moment $S = 1/2$ and effective angular moment $L_{eff} = 1$, splits into the ground state $J_{eff} = 1/2$ Kramers doublet and the higher $J_{eff} = 3/2$ quartet at $3\lambda_{SO}/2$ due to the spin-orbit coupling between $S$ and $L_{eff}$. This insulating state with $J_{eff} = 1/2$ moment determined from the *LS*-coupling scheme is essentially equivalent to the $j_{eff} = 1/2$ Mott insulator, since both have only one-hole involved in the ground state. The difference between the two limits appears when the $Ir^{5+}$, $d^4$ configuration with unquenched Hund's coupling, is generated by virtual charge fluctuations. In this case, as discussed below, the Hund's coupling between two holes in the virtual $d^4$ configuration is essential for a correct description of the exchange interactions and the *LS*-coupling picture might be more convenient. In the other part of this review, we use capital $J_{eff}$ for simplicity.

In the case of the $Ru^{3+}$ chloride with five $4d$ electrons, the same picture as the $Ir^{4+}$ oxides can be applied. As the energy scale of Hund's coupling is not very different from that of $\lambda_{SO}$ for $Ir^{4+}$, the real situation in iridates is highly likely in between the *jj*- and the *LS*-coupling limits. As a smaller $\lambda_{SO} \sim 0.1$ eV and a larger Hund's coupling of ~0.4 eV are expected for the $Ru^{3+}$ ion with five $4d$ electrons, the real situation for the Ru chloride should be closer to the *LS*-coupling scheme than the $Ir^{4+}$ oxides.

In the Ir$^{4+}$ oxides and the Ru$^{3+}$ chlorides, the spin-orbit coupling is much larger than the typical exchange interactions among magnetic ions. The low-energy magnetism of these compounds is thus dominated by the $J_{eff}$ = 1/2 moments. The spin-orbital entangled $J_{eff}$ = 1/2 wave function is composed of the quantum superposition of $d_{xy}$, $d_{yz}$, and $d_{zx}$ orbitals with equal amplitudes but complex phases that describe orbital motion. The up- and the down-spins, ↑ and ↓, reside on different orbital states:

$$|J_{eff}=\tfrac{1}{2}, J_{eff}^z=+\tfrac{1}{2}\rangle = \tfrac{1}{\sqrt{3}}(|d_{xy},\uparrow\rangle + |d_{yz},\downarrow\rangle + i|d_{zx},\downarrow\rangle).$$

$$|J_{eff}=\tfrac{1}{2}, J_{eff}^z=-\tfrac{1}{2}\rangle = -\tfrac{1}{\sqrt{3}}(|d_{xy},\downarrow\rangle - |d_{yz},\uparrow\rangle + i|d_{zx},\uparrow\rangle).$$

The $J_{eff}$ = 1/2 state has electron density distribution of cubic shape, as shown in **Fig. 2b**, and hosts a magnetic moment of 1 $\mu_B$ exactly like a free spin $S$ = 1/2. However, its gyromagnetic factor $g$ = -2 is opposite to that of spin, which is a manifestation of the unquenched orbital moment $L$. Via the orbital component $L$, $J_{eff}$ = 1/2 moments and their exchange interactions are extremely sensitive to the local crystalline fields and the bonding geometry[23,37].

When neighboring IrO$_6$ octahedra share one of their corner oxygens to form 180° Ir-O-Ir bonds, the super-exchange interaction between the two adjacent $J_{eff}$ = 1/2 moments on Ir$^{4+}$ is dominated by the isotropic Heisenberg term, despite the strong spin-orbit coupling[22,38,39]. The emergence of isotropy (pseudo-spin rotational symmetry) occurs because the electron hopping between neighboring ions conserves not only the spin but also the orbital index, and therefore the $J_{eff}$ = 1/2 quantum number. On the other hand, when the two IrO$_6$ octahedra share one of their edges to form 90° Ir-O$_2$-Ir bonds, as shown in **Fig. 2c**, the oxygen-mediated hopping becomes orbital non-conserving, leading to exchange couplings with a discrete symmetry. A destructive quantum interference of the two Ir-O-Ir paths in the super-exchange process completely suppresses the conventional Heisenberg term between the neighboring $J_{eff}$ = 1/2 moments. Instead, an Ising ferromagnetic exchange $-KJ^{\gamma}_{eff,i}J^{\gamma}_{eff,j}$, with easy axes perpendicular to the Ir-O$_2$-Ir plane, emerges via a combination of hopping to the neighboring $J_{eff}$ = 3/2 and the subsequent Hund's coupling, which favors parallel alignment of the real spins (**Fig. 2d**)[23]. By replacing the spin-orbital entangled pseudo-spin $J_{eff}$ with $S$, one arrives at the essential ingredients of the Kitaev model: bond-specific $S$ = 1/2 Ising interactions $K_\gamma S^\gamma_i S^\gamma_j$ (Box 1) with $\gamma$-axis perpendicular to the Ir-O$_2$-Ir bond plane (**Fig. 3a**).

$A_2$IrO$_3$ ($A$ = Na, Li) was first proposed as a playground for the materialization of the Kitaev model[23], and consists of alternating two-dimensional layers of IrO$_6$ octahedra forming a honeycomb network (**Fig. 3b**) and Na (Li) as shown in **Fig. 3c**. See **Box 2** for the detailed description of the structure of Na$_2$IrO$_3$. Each IrO$_6$ octahedron shares its edges with the neighboring IrO$_6$ octahedra and forms three 90° Ir-O$_2$-Ir bonds, with the bonding plane orthogonal to the other two as shown in **Fig. 3a**. The super-exchange process through Ir-O$_2$-Ir bonds gives rise to three kinds of Ising ferromagnetic bonds, with the bond-dependent easy-axes orthogonal to each other. If the interaction between the neighboring pseudo-spin $J_{eff}$ = 1/2 moments is dominated by such bond-dependent Ising interactions, the honeycomb network of

$A_2IrO_3$ is nothing but *a materialization of the Kitaev model*. The marriage of QSL physics and correlated electron physics kicked off the exploration of the Kitaev candidate materials including $Na_2IrO_3$[40]. As discussed below, additional interactions always exist in materials, and cannot be neglected.

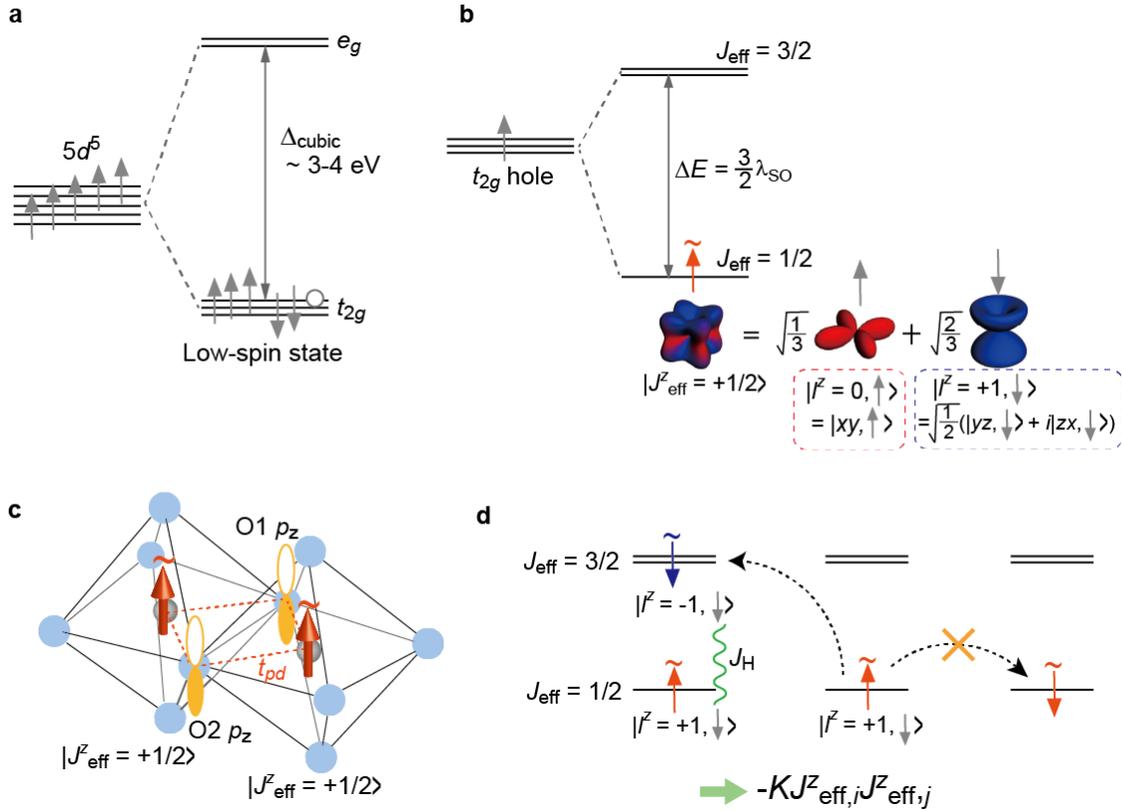

**Figure 2. Local electronic states of octahedrally coordinated $Ir^{4+}$ and $Ru^{3+}$ ions. a**, Splitting of five-hold degenerate $d$-levels of $d^5$ $Ir^{4+}$ and $Ru^{3+}$ into the doublet $e_g$ and the triplet $t_{2g}$ due to the cubic crystal field $\Delta_{cubic}$ of $O_6(Cl_6)$ octahedron. **b**, Spin-orbit coupling $\lambda_{SO}$ further splits the $t_{2g}$ manifold into the $J_{eff} = 1/2$ doublet and $J_{eff} = 3/2$ quartet by $\Delta E = 3\lambda_{SO}/2$. One hole is accommodated in the lower $J_{eff} = 1/2$ Kramers doublet in the hole picture. **c**, The edge-shared bond between the two $IrO_6$ ($RuCl_6$) octahedra gives rise to two 90° Ir-O-Ir (Ru-Cl-Ru) bonds. Hopping of the hole via the two oxygen $p_z$ orbitals ($t_{pd}$) changes the orbital "color", i.e. $zx$ to $yz$ and $yz$ to $zx$ and a quantum interference between the two hopping paths gives a selection rule $\Delta l^z = \pm 2$. **d**, As a result the hopping between the neighboring $J_{eff} = 1/2$ orbitals are prohibited and the conventional Heisenberg exchange is suppressed. Instead, the hopping between the neighboring $J_{eff} = 1/2$ and $J_{eff} = 3/2$ (more specifically $J^z_{eff} = \pm 3/2$) orbitals dominates, and the Hund's coupling $J_H$ between real spins indicated by green wavy line leads to anisotropic Ising ferromagnetic coupling. The "Isingness" of this process follows from the hopping selection rules $\Delta l^z = \pm 2$ and $\Delta s^z = 0$, which dictate that a hole must return to the $J_{eff} = 1/2$ level with the same quantum numbers.

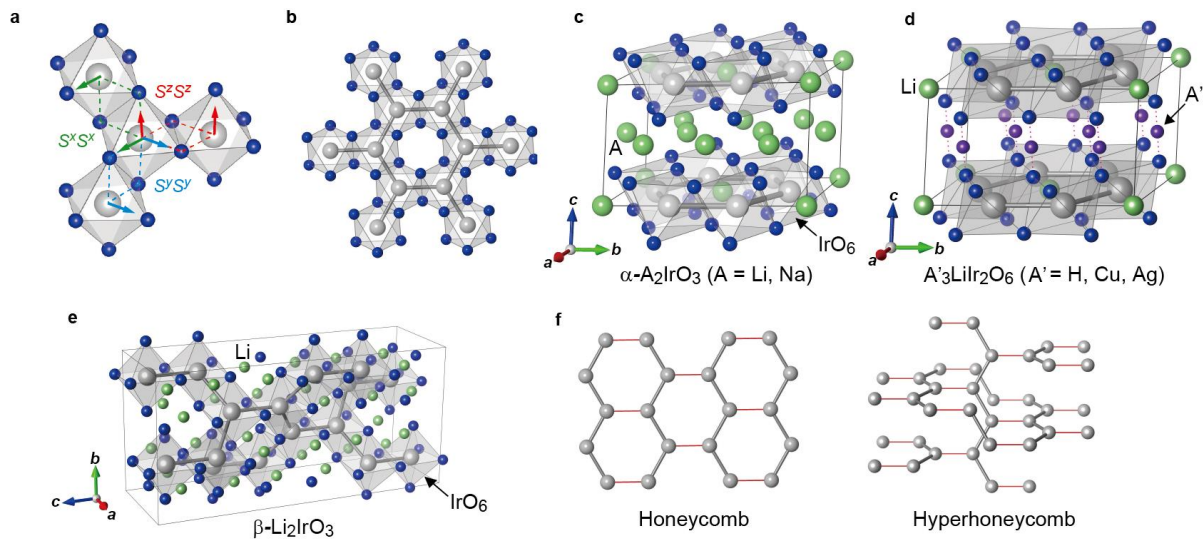

**Figure 3. Crystal structures of Kitaev materials**. **a,** The edge-shared bonds with the neighboring three octahedra, accommodating Kitaev-type interactions. **b,** Honeycomb network of $IrO_6$ ($RuCl_6$) octahedra, commonly seen in $Na_2IrO_3$, $\alpha$-$Li_2IrO_3$, $\alpha$-$RuCl_3$ and the other two-dimensional honeycomb iridates in Table 1. **c,** The crystal structure of $Na_2IrO_3$ and $\alpha$-$Li_2IrO_3$. **d,** The crystal structure of A'$_3$LiIr$_2$O$_6$ (A' = H, Cu and Ag). **e,** The crystal structure of $\beta$-$Li_2IrO_3$ with hyperhoneycomb Ir sublattice shown in **f**. **f,** The relationship between the two-dimensional honeycomb and the three-dimensional hyperhoneycomb lattices. The bonds in honeycomb lattice are decomposed into the zigzag chains (gray) and bridges (red). The crystal structures are visualized by using VESTA software[130].

**Kitaev candidate materials**

After the theoretical prediction of Kitaev physics in Na$_2$IrO$_3$[23], many iridium and ruthenium compounds were recognized as candidates to materialize the Kitaev model, as summarized in Table 1. The Na$^+$ ions in Na$_2$IrO$_3$ can be replaced with Li$^+$ and Cu$^+$ ions[41, 42]. $\alpha$-Li$_2$IrO$_3$ has the same structure as Na$_2$IrO$_3$ (**Fig. 3c**), and was investigated previously as a possible material for Li-ion batteries[43] and rediscovered as a Kitaev candidate[44]. Applying an ion-exchange technique to $\alpha$-Li$_2$IrO$_3$, one can replace the Li$^+$ ions sitting in-between the honeycomb layers with other monovalent ions such as H$^+$, Cu$^+$ and Ag$^+$. This gives rise to a second generation of two-dimensional Kitaev candidates, H$_3$LiIr$_2$O$_6$, Cu$_3$LiIr$_2$O$_6$ and Ag$_3$LiIr$_2$O$_6$[45-49]. The chemical formula A'$_3$LiIr$_2$O$_6$ (A' = H, Cu and Ag) can be compared with $\alpha$-Li$_2$IrO$_3$ = 1/2 (Li'$_3$LiIr$_2$O$_6$), where Li' represents Li ions between the honeycomb LiIr$_2$O$_6$ layers. The interlayer A' ions in A'$_3$LiIr$_2$O$_6$ form a straight dumbbell bond with oxygens above and below, resulting in a different stacking pattern of LiIr$_2$O$_6$ layers from those of Na$_2$IrO$_3$ and $\alpha$-Li$_2$IrO$_3$[47] (see **Fig. 3d**). $\alpha$-RuCl$_3$ comprises essentially the same honeycomb network of edge-sharing RuCl$_6$ octahedra and can be viewed as fully Na-deficient analogue of Na$_2$IrO$_3$[50,51].

The polytypes $\beta$-Li$_2$IrO$_3$[52] (**Fig. 3e**) and $\gamma$-Li$_2$IrO$_3$[53] were discovered as three-dimensional analogues of two-dimensional honeycomb $\alpha$-Li$_2$IrO$_3$. The two-dimensional honeycomb lattice can be viewed as an assembly of one-dimensional zigzag chains connected by the bridging bonds, all confined in the same plane. Consider the three-dimensional stack of such two-dimensional honeycomb plane. In the Ir-sublattice of $\beta$-Li$_2$IrO$_3$, called hyperhoneycomb, the zigzag chains in the three-dimensional stack are rotated around the bridging bonds in an alternating pattern and reconnected to the zigzag chains in the layers above and below by the bridging bonds[52] as depicted in **Fig. 3f**. In $\gamma$-Li$_2$IrO$_3$, called stripy-honeycomb, the stripes with hexagons, consisting of two zigzag chains with bridging bonds in between, are rotated alternately around the stripe-bridging bonds and reconnected to the stripes in the layers above and below[53]. In both $\beta$-Li$_2$IrO$_3$ and $\gamma$-Li$_2$IrO$_3$, all the Ir sites are crystallographically equivalent and remain octahedrally coordinated by oxygen atoms. As in $\alpha$-Li$_2$IrO$_3$, each IrO$_6$ octahedron shares its three orthogonal edges with the three neighboring IrO$_6$ octahedra, forming three bonds. The local bonding configurations of $\beta$-Li$_2$IrO$_3$ and $\gamma$-Li$_2$IrO$_3$ are the same as $\alpha$-Li$_2$IrO$_3$ and should give rise to the same competition between three Ising bonds with orthogonal easy axes. The three-dimensional Kitaev model on the hyperhoneycomb ($\beta$-Li$_2$IrO$_3$-type) lattice was investigated even before the compounds were discovered and shown to have a QSL as its ground state[54]. These 3D honeycomb iridates represent the materialization of such three-dimensional Kitaev models.

It should be emphasized that in all the candidates the IrO$_6$(RuCl$_6$) octahedra are not regular but have small non-cubic distortions[41,47,51,55]. In case of the two-dimensional honeycomb compounds, the two O$_3$ (Cl$_3$) triangles above and below Ir (Ru), facing each other and forming the O$_6$(Cl$_6$) octahedron, come closer to each other. As a result of this compression perpendicular to the layer direction, the Ir-O-Ir (Ru-Cl-Ru) angle becomes more than 90° and a trigonal crystal field is produced. This should modify the bond-sensitive exchange interactions, very likely modifying the Kitaev-type interaction and enhancing some of the other interactions. We also note that the interlayer coupling of those layered compounds is in general very weak.

All the samples reported so far suffer from the presence of stacking faults which manifests as a broadening of x-ray diffraction peaks[47,48]. In case of $\alpha$-RuCl$_3$, the sample dependence of the refined crystal structures and, as discussed below, the magnetic ordering transitions were reported[51].

**Electronic structure of Kitaev candidates**

All the Kitaev candidate materials in Table 1 exhibit an insulating behavior and a localized moment magnetism, indicating that they are $J_{eff} = 1/2$ spin-orbital Mott insulators. The optical conductivity spectrum for Na$_2$IrO$_3$ shows a sizable charge gap of ~300 meV[56], which is consistent with an activation energy of ~100 meV at room temperature in the resistivity, $\rho(T)$[40,57]. Transport activation energies comparable to that of Na$_2$IrO$_3$ have been observed in the other honeycomb-based iridium oxides[44,52], implying similar charge gaps. The relatively small size of the charge gap suggests that they form weak Mott insulators due to the moderate Coulomb $U$ ~ 2 eV. For $\alpha$-RuCl$_3$, optical conductivity measurements show a charge gap of ~1 eV, substantially larger than that of Na$_2$IrO$_3$, likely reflecting a larger $U$ for 4$d$ Ru and a narrow band width originating from the strong ionicity of chloride[58].

In the resonant inelastic x-ray scattering (RIXS) spectra at the Ir $L_3$ ($2p_{3/2} \rightarrow 5d$) edge for Na$_2$IrO$_3$, $\alpha$-Li$_2$IrO$_3$ and $\beta$-Li$_2$IrO$_3$, the presence of low-energy excitations of ~0.7 eV is commonly observed[59,60], which corresponds to the excitation energy of $3\lambda_{SO}/2$ from $J_{eff} = 1/2$ to 3/2 within the $t_{2g}$ manifold, and indicates a spin-orbit coupling $\lambda_{SO}$ ~ 0.4-0.5 eV. A splitting of the 0.7 eV peak is observed[59], which likely originates from the splitting of $J_{eff} = 3/2$ states due to the trigonal lattice distortion of IrO$_6$ octahedron mentioned above. The splitting is smaller than $3\lambda_{SO}/2$ ~ 0.7 eV, meaning that the $J_{eff} = 1/2$ picture is valid as a first approximation, but the effect of trigonal crystal field may not be negligibly small.

The presence of excitations between $J_{eff} = 1/2$ and 3/2, analogues to the iridates, was identified by INS in $\alpha$-RuCl$_3$[61], and is consistent with the expected $\lambda_{SO}$ of 0.1 eV for 4$d$ Ru. Despite the small spin-orbit coupling, the $J_{eff} = 1/2$ picture holds for $\alpha$-RuCl$_3$, and is likely an even better approximation than that in the iridium oxides. The RuCl$_6$ octahedron is less trigonally distorted than the IrO$_6$ octahedra in the iridium oxides, which reduces the crystal-field splitting to a smaller value than the spin-orbit coupling of $\lambda_{SO}$ ~ 0.1 eV[62].

**Magnetism of Kitaev candidates**

A numerical calculation of magnetic susceptibility $\chi(T)$ for the Kitaev model with the uniform ferromagnetic couplings $K = K_x = K_y = K_z$ (see **Box 1** for the Kitaev model) indicates an isotropic Curie-Weiss behavior with a Curie constant for $S = 1/2$ and $g = 2$, and a ferromagnetic Curie-Weiss temperature $\theta_{CW} = K/4$ at high temperatures. Upon lowering temperature $T$, $\chi(T)$ shows a downward deviation from the Curie-Weiss behavior below $T \sim K/k_B$ and then crosses over to almost $T$-independent behavior around a characteristic temperature $T_H$ below which the spin-spin correlations saturate[63].

**Figures 4a-d** summarize the magnetic susceptibilities $\chi(T)$ of the Kitaev candidate compounds[40,46,51,52,64]. The relevant magnetic parameters estimated from $\chi(T)$ in **Fig. 4** are listed in **Table 1**. The Curie-Weiss behavior at high temperatures is indicative of the localized moment magnetism and is apparent from the almost $T$-linear behavior of the inverse susceptibility $1/\chi(T)$ in the Curie-Weiss plots in **Figs. 4b, c and d**. The effective moments, determined from the slope of the linear behavior, are close to $p_{eff}$ = 1.73 $\mu_B$ expected for the pure $J_{eff}$ = 1/2 state, and equivalent to the case for $S$ = 1/2 and $g$ = 2, and are material independent. This is consistent with $J_{eff}$ = 1/2 Mottness of all the compounds in **Table 1**. The presence of a sizable field-orientation-dependent anisotropy in $\chi(T)$ suggests the presence of bond-dependent anisotropic magnetic couplings and the trigonal field effect. The deviation from the Curie-Weiss behavior can be seen at temperatures between 100 K and 200K, indicating a rough energy scale of magnetic interactions of 100 - 200 K. The Curie-Weiss temperatures $\theta_{CW}$ are negative (antiferromagnetic), except for $\alpha$-RuCl$_3$ ($B$ // plane) and $\beta$-Li$_2$IrO$_3$ ($B$ // $b$- or $c$-axis), and depend strongly on the materials and the field orientations, ranging from -200 K (antiferromagnetic) to almost zero. It is clear that the Kitaev-type ferromagnetic coupling is not the only important interaction and that appreciable antiferromagnetic interactions are present. The almost zero $\theta_{CW}$ despite the deviation from the Curie-Weiss behavior above 100 K indicates a cancellation of antiferromagnetic and ferromagnetic interactions, evidencing the presence of more than one kind of interactions in these compounds. The wide scattering of $\theta_{CW}$ very likely mirrors the sensitivity of the interactions to the local lattice structure and the details of chemical bonding, arising from the spin-orbital entanglement.

Except for H$_3$LiIr$_2$O$_6$, all the candidate compounds show a clear signature of long-range magnetic ordering rather than a QSL state, which is evident from the kink in $\chi(T)$ and the specific heat anomaly in $C(T)$ shown in **Fig. 4e**. The ordering temperature $T_{mag}$ is one order of magnitude lower than the energy scale of magnetic interaction of 100 - 200 K inferred from the deviation from Curie-Weiss behavior. It indicates the presence of magnetic frustration, which is consistent with their possible proximity to the frustrated Kitaev QSL. In accord with the presence of frustration, the entropy change around $T_{mag}$ is only 10-30% of $R$ln2[40,51,52,65], the full entropy of $J_{eff}$ = 1/2 doublet.

Resonant x-ray diffraction and neutron measurements indicate that Na$_2$IrO$_3$[66,67] and $\alpha$-RuCl$_3$[51,68,69] form a zigzag ordering below $T_{mag}$, where ferromagnetic zigzag chains are coupled antiferromagnetically across the bridging bonds. $\alpha$-Li$_2$IrO$_3$ shows a spiral ordering below $T_{mag}$ along the zigzag chains[70]. The two distinct ordering patterns are depicted in **Fig. 4f**. The three-dimensional honeycomb iridates, $\beta$-Li$_2$IrO$_3$ and $\gamma$-Li$_2$IrO$_3$ show a complex incommensurate spiral order below $T_{mag}$ ~ 38 K[71,72]. The parallel behavior of polymorphic $\beta$-Li$_2$IrO$_3$ and $\gamma$-Li$_2$IrO$_3$, including indistinguishably close magnetic ordering temperatures, has been suggested to be due to the identical local connectivity of the three Ir-bonds[73]. The details of the magnetic ordering transition in $\alpha$-RuCl$_3$ are sample dependent: crystals with minimal stacking faults generally show a sharp transition at $T_{mag} \approx$ 7 K[51], corresponding to zigzag order in the plane with a three-layer periodicity. Samples with significant presence of stacking faults show an additional broad transition near 14 K corresponding to a two-layer periodicity of the zig-zag structure, and powders show only the broad 14 K transition. It is possible that other periodic layerings exist corresponding to different values of $T_{mag}$.

The emergence of long-range magnetic ordering rather than a QSL state is believed to originate from the presence of additional interactions beyond the Kitaev model. While the presence of dominant Kitaev interactions is shown for example from the azimuthal-angle dependence of resonant x-ray diffuse magnetic scattering in $Na_2IrO_3$[74] and the dispersion of spin excitations obtained from INS in $\alpha$-$RuCl_3$[61,75], there are other types of exchange interactions such as Heisenberg exchange through direct overlap of $d$-orbitals[76] and off-diagonal exchanges like the $S_i^x S_j^y$ term[77-79], which bring the real materials away from the pure Kitaev model regime. Lattice distortions, such as compression of the $IrO_6$ octahedron perpendicular to the plane, give rise to an increased Ir-O-Ir bond angle from 90° and admixture of $J_{eff} = 3/2$ states with $J_{eff} = 1/2$ states[78,79], which may add extra exchange paths[80]. Further-neighbor interactions could also be relevant[44,55,81-83]. The Kitaev-Heisenberg model with additional nearest-neighbor Heisenberg interactions was discussed first, and gives a Kitaev spin liquid in the pure Kitaev limit, Neel order in the strong Heisenberg limit, and stripy order in-between[76]. The experimentally observed zigzag and spiral orderings are not contained in the theoretical phase diagram of the ferromagnetic Kitaev-Heisenberg model. This led to the discussion of the off-diagonal and further-neighbor interactions, which for a certain range of parameters reproduce the zigzag and incommensurate spiral orderings[77-80,84,85]. (See **Box 3** for details.)

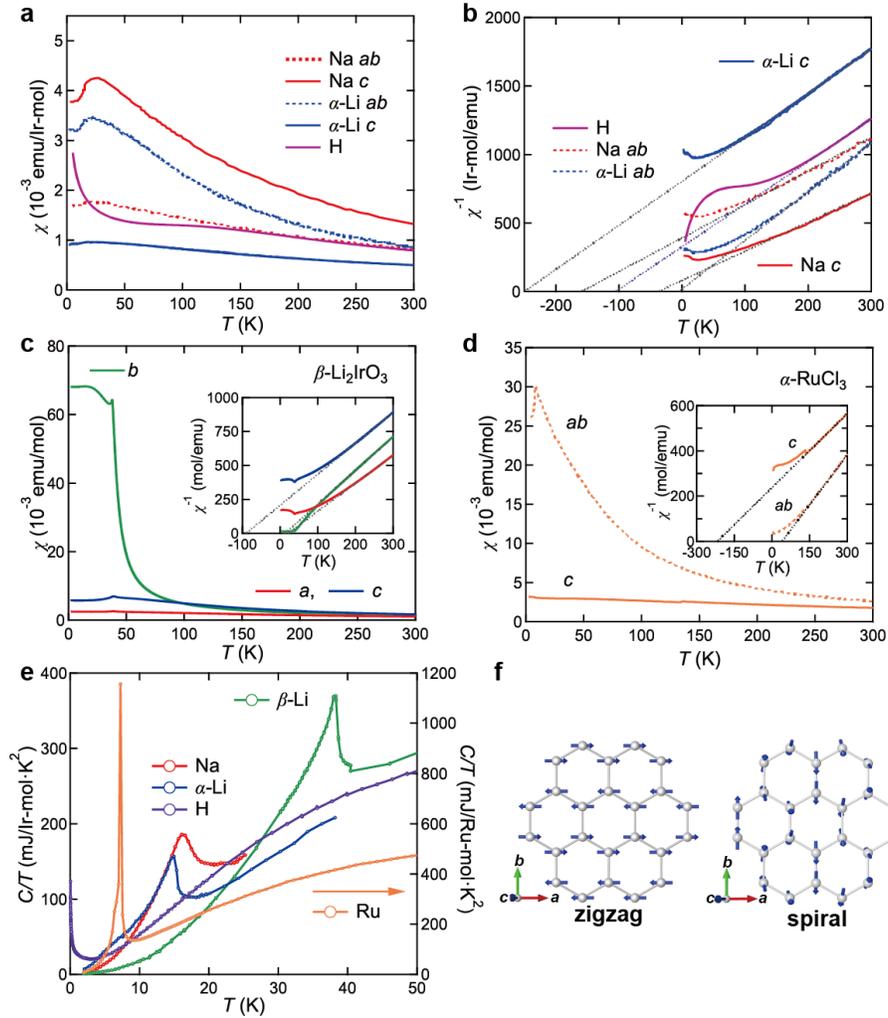

**Figure 4**. **Magnetic and thermodynamic properties of Kitaev candidate materials. a,** Magnetic susceptibility $\chi(T)$ for two-dimensional honeycomb iridate $Na_2IrO_3$ (Na)[40], $\alpha$-$Li_2IrO_3$ ($\alpha$-Li) and $H_3LiIr_2O_6$ (H)[46] with magnetic fields parallel ($ab$) and perpendicular ($c$) to the honeycomb plane. **b,** The Curie-Weiss plot, $\chi^{-1}(T)$ as a function of $T$, of data in **a**. **c,** $\chi(T)$ for three-dimensional hyperhoneycomb iridate $\beta$-$Li_2IrO_3$ with magnetic fields parallel $a$-, $b$- and $c$- axes. **d,** $\chi(T)$ for two-dimensional honeycomb ruthenium compound $\alpha$-$RuCl_3$ with magnetic fields parallel ($ab$) and perpendicular ($c$) to the honeycomb plane[64]. The inset shows the Curie-Weiss plot of data in the main panel. **e,** The low temperature specific heat $C(T)$ for $Na_2IrO_3$ (Na), $\alpha$-$Li_2IrO_3$ ($\alpha$-Li)[65], $H_3LiIr_2O_6$ (H)[46], $\alpha$-$RuCl_3$ (Ru)[51] and $\beta$-$Li_2IrO_3$ ($\beta$-Li). **f,** Zigzag and spiral ordering patterns for $Na_2IrO_3$ and $\alpha$-$Li_2IrO_3$.

**Table 1. Representative Kitaev candidate materials and summary of their physical properties.**

(*$R$-$3m$ assumed in Ref.[48] because of strong stacking disorder)

| Materials | Crystal structure (Space group) | $T_{mag}$ | anisotropy | $p_{eff}$ ($\mu_B$) | $\theta_{CW}$ (K) | Magnetic ground state | Ref. |
|---|---|---|---|---|---|---|---|
| $Na_2IrO_3$ | 2D ($C2/m$) | 15 K | $\chi_c > \chi_{ab}$ | 1.81 ($ab$) 1.94 ($c$) | -176 ($\theta_{ab}$) -40 ($\theta_c$) | zigzag | 40,57,66, 67 |
| $\alpha$-$Li_2IrO_3$ | 2D ($C2/m$) | 15 K | $\chi_{ab} > \chi_c$ | 1.50 ($ab$) 1.58 ($c$) | +5 ($\theta_{ab}$), -250 ($\theta_c$) | Spiral | 44,65,70 |
| $H_3LiIr_2O_6$ | 2D ($C2/m$) | - | $\chi_{ab} > \chi_c$ | 1.60 | -105 | Spin-liquid | 46 |
| $Cu_2IrO_3$ | 2D ($C2/c$) | 2.7 K | Not known | 1.93(1) | -110 | AF order or Spin-glass | 42 |
| $Cu_3LiIr_2O_6$ | 2D ($C2/c$) | 15 K | Not known | 2.1(1) | -145 | AF order | 49 |
| $Ag_3LiIr_2O_6$ | 2D ($R$-$3m$*) | ~12 K | Not known | 1.77 | | AF order | 48 |
| $\alpha$-$RuCl_3$ | 2D ($C2/m$ or $P3_112$, or $R$-$3$); $T$ and sample dependent | 7 K and/or, 14 K See text | $\chi_{ab} > \chi_c$ | 2.33 ($ab$), 2.71 ($c$) | +39.6($\theta_{ab}$), -216.4 ($\theta_c$) | zigzag | 51,64,68, 69, 131 |
| $\beta$-$Li_2IrO_3$ | 3D ($Fddd$) | 38 K | $\chi_b > \chi_c > \chi_a$ | 1.87 ($a$) 1.80 ($b$) 1.97 ($c$) | -90.2 ($\theta_a$) +12.9 ($\theta_b$) +21.6 ($\theta_c$) | Spiral | 52,71,92 |
| $\gamma$-$Li_2IrO_3$ | 3D ($Cccm$) | 39.5 K | $\chi_b > \chi_c > \chi_a$ | ~1.6 | +40 | Spiral | 53,72 |

## Phase control of Kitaev candidate materials

Despite the emergence of magnetic order at low temperatures, the candidate materials for the Kitaev spin liquid seem to have sizable Kitaev interactions and may be tunable into a spin-liquid ground state by external perturbations. The results of numerical calculations for an extended Kitaev model with Heisenberg and off-diagonal interactions (See **Box 3**) indeed indicates the presence of a QSL for a finite region of parameter space[77-79]. The suppression of magnetic ordering has been attempted by applying magnetic field or high pressure.

$\alpha$-RuCl$_3$ undergoes a zigzag-type magnetic ordering at $T_{mag} \sim 7$ K, which was found to be suppressed drastically by applying an in-plane magnetic field of $B_c \sim 7 - 8$ T[69,86,87], as shown in Figs. 5**a** and **b**. Above $B_c$, the system appears to show no magnetic ordering down to well below 1 K. Around $B_c$, an induced moment of 0.6 $\mu_B$ is observed[69] meaning a substantial portion of magnetic entropy must be lifted. Nevertheless, unusual magnetic excitations and thermal transport were discovered in the field-induced critical paramagnet[88-91], which were discussed in connection with a Kitaev QSL and are attracting considerable attention from the community. This will be described in detail in a later section.

Similar $B$-induced suppression of the magnetic order is observed in three-dimensional honeycomb $\beta$-Li$_2$IrO$_3$ and $\gamma$-Li$_2$IrO$_3$. By applying a magnetic field along the $b$-axis, which is the expected Ising axis of bridging bonds, the magnetic transition is smeared out and disappears above $B_c \sim 3$ T, where the magnetic moment of 0.4 $\mu_B$ is observed[52,53,92,93]. This field-induced paramagnetic phase of three-dimensional honeycomb compounds is much less explored than that of $\alpha$-RuCl$_3$.

Pressure is another external perturbation to control the magnetic interactions and thus the magnetic ground state. In $\alpha$-Li$_2$IrO$_3$, $\alpha$-RuCl$_3$ and three-dimensional $\beta$-Li$_2$IrO$_3$, a first-order transition to a dimerized state is commonly observed[60,94-98], where a modulation of zigzag chains, the alternating contraction and elongation of Ir-Ir (Ru-Ru) bonds, takes place. The Ir-Ir distance of contracted bonds is even shorter than that of metallic Ir, suggesting the formation of strong Ir$_2$-dimer molecules rather than a weak spin-Peierls-like transition[60,95,97]. Similar strong dimerization occurs in many 3$d$ and 4$d$ transition metal oxides including Li$_2$RuO$_3$ with the same honeycomb structure[99]. In the $d^5$ iridium and ruthenium compounds, the strong dimerization appears to compete with the $J_{eff} = 1/2$ spin-orbital Mott state with isotropic superposition of the three orbital states, as a specific orbital state forming the bond is selected in the dimerized state.

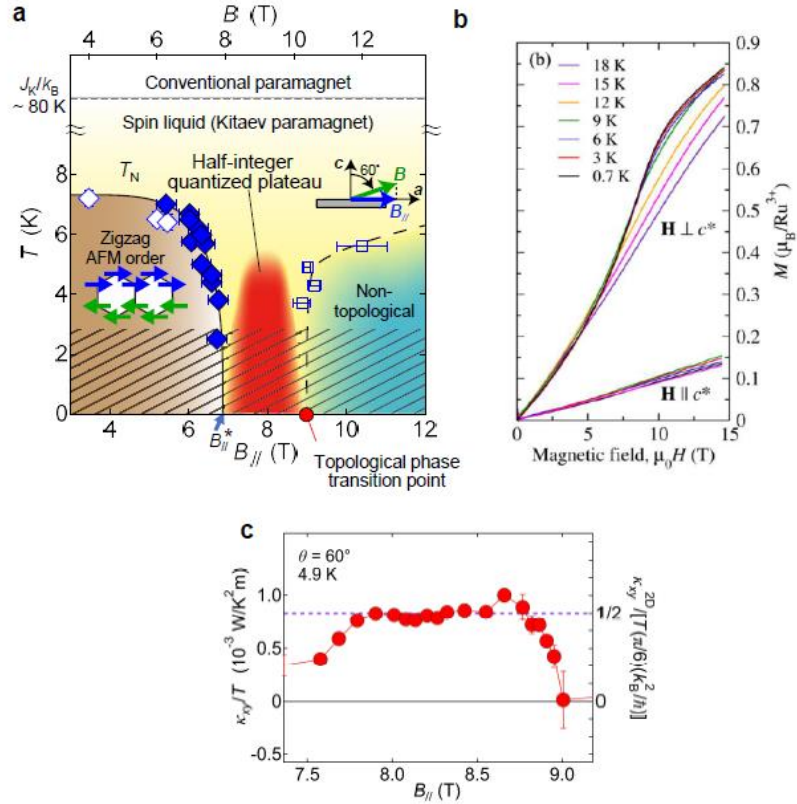

**Figure 5. Magnetic field induced collapse of zigzag magnetic ordering in $\alpha$-RuCl$_3$. a,** Proposed phase diagram of $\alpha$-RuCl$_3$ as a function of the in-plane magnetic field $B_{//}$ [91]. **b.** Magnetization curve of $\alpha$-RuCl$_3$ with in-plane ($B \perp c^*$) and out-of-plane ($B // c^*$) magnetic fields[69]. **c,** Quantized thermal Hall effect $\kappa_{xy}$ as a function of the $B_{//}$ measured under 60° tilted magnetic field. The dotted line indicates a half-quantized value $\kappa_{xy}/T = 1/2(\pi/6)(k_B^2/\hbar)$[91]. (Panel **a** and **c** adapted with permission from Ref. 91, and panel **b** reproduced with permission from Ref. 69).

## Quantum spin liquid state in $H_3LiIr_2O_6$

Another approach to control the magnetic interactions is chemical modification. The second generation of two-dimensional honeycomb iridates, $A'_3LiIr_2O_6$ (A' = H, Cu and Ag)[45-49], may be the typical playground for such an approach. A QSL state was recently discovered in "protonated" $H_3LiIr_2O_6$[46]. The honeycomb layers of $H_3LiIr_2O_6$ are more compressed perpendicular to the plane than the first generation of two-dimensional honeycomb iridates[45,47], which could modify the magnetic interactions appreciably. The evidence for the absence of (short- and long-ranged) magnetic ordering can be seen in the $^7$Li- and $^1$H-NMR spectra shown in **Fig. 6**, which do not show any broadening down to below 1 K. The Knight shift $K_s(T)$ obtained from the spectra in **Fig. 6c,** as well as $\chi(T)$ in **Fig. 4b,** shows a Curie-Weiss behavior with $\theta_{CW}$ ~ -100 K at high temperatures, analogous to those of the first-generation candidates. With lowering temperature below 150 K, $K_s(T)$ of $H_3LiIr_2O_6$ shows a broad peak around 130 K, in contrast to the first generation, and decreases to a finite value in the $T = 0$ limit. Signatures of local, energy-symmetric, low-energy fermionic excitations are observed in the magnetization $M(T, B)$, NMR relaxation $1/T_1(T, B)$ and specific heat $C(T, B)$ at low temperatures, originating from a small number of spin defects. The defect contributions in $M$, $1/T_1$ and $C$ follow a scaling with $B/T$, and, while no theoretical consensus has emerged, it has been suggested that this could arise from random singlet formation of spin defects embedded in the bulk QSL[100] or from a peculiar band dispersion of Majorana fermions produced by an interlayer coupling[101]. After subtracting the scaled contribution originating from the defects, only a $B$-independent $T^3$-contribution to $C(T)$ is observed below $T = 5$ K, which very likely originates from the lattice. The absence of appreciable magnetic entropy at low temperatures may suggest the presence of a gap in the spin excitations. The observed behavior is distinct from that expected for the "pure" Kitaev QSL, which has a specific heat anomaly with entropy of 50% of $R$ln2 associated with ordering the $Z_2$ fluxes (localized Majorana fermions)[32]. It was argued that the presence of non-Kitaev interactions may change the nature of the QSL appreciably from that in the pure Kitaev limit[102]. Also for $H_3LiIr_2O_6$, the randomness of the H positions was discussed as playing an important role in stabilizing the QSL state[103, 104]. Unveiling the connection/disconnection to the Kitaev physics is an interesting challenge for the near future.

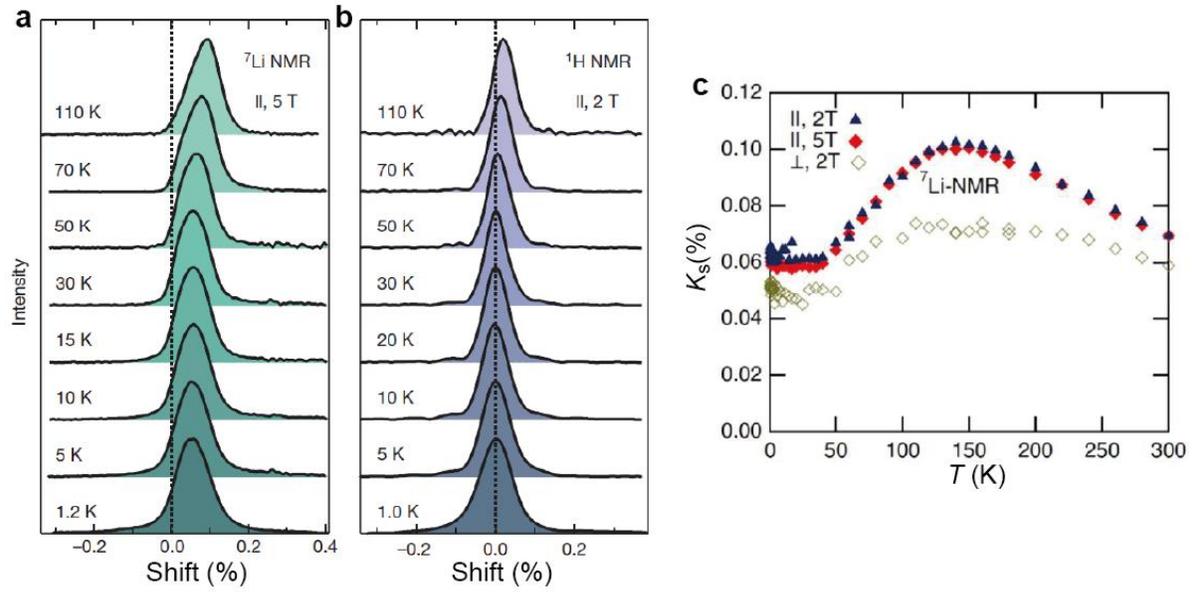

**Fig. 6 Quantum spin liquid state in $H_3LiIr_2O_6$[46]. a, b**, $^7$Li- and $^1$H-NMR spectra at various temperatures, showing no broadening down to a low temperature. **c**, temperature dependent Knight shift $K_s(T)$ with magnetic field parallel (//) and perpendicular ($\perp$) to the honeycomb plane. (Panel **a** and **b** adapted with permission from Ref. 46).

**Magnetic excitations and evidence of fractionalization in Kitaev candidate materials**

As discussed above, one defining characteristic of the QSL is the presence of fractional magnetic excitations, and the predicted Majorana fermions in the Kitaev QSL are of widespread interest. A wide variety of spectroscopic techniques have been applied to examine magnetic excitations in iridates, and in particular $\alpha$-RuCl$_3$, including INS[55,61,75,90,105,106], Raman spectroscopy[107,108], ESR[109], NMR[88,110,111], and THz spectroscopy[112-115].

With the exception of H$_3$LiIr$_2$O$_6$[46], the Kitaev candidate materials discussed here order magnetically at low temperature, albeit with reduced ordered moments. In principle the magnetically ordered states can support conventional spin wave modes. In collinear magnets without frustrated or competing interactions a single-crystal measurement of the spin-wave dispersions and intensities via INS often enables the direct determination of the effective spin Hamiltonian parameters. The situation in the honeycomb magnets with possible Kitaev interactions is more complicated. The spin-wave spectra themselves can be readily calculated using parameters proposed to describe $\alpha$-RuCl$_3$ from theoretical considerations[116,117]. There is, however, more than one set of parameters that can reproduce the experimental results reasonably, likely because of strong frustration. The iridate materials present additional challenges since Ir is a strong neutron absorber and the magnetic form factor appearing in the neutron scattering cross section[118] decays rapidly as the magnitude of the momentum transfer is increased. Moreover, single crystals suitable for INS are not necessarily available, forcing measurements to be made using polycrystalline material. Despite these difficulties, INS measurements of Na$_2$IrO$_3$ powder[55] proved to be very informative. For example, the shape of the scattering threshold alone established the magnetic order as zigzag, not stripy (see **Box 4**). Similar measurements on Ru based materials such as $\alpha$-RuCl$_3$ are generally much easier. In $\alpha$-RuCl$_3$ powders the shape of the low-energy, low-momentum scattering threshold is well-defined and an additional high energy feature is detected with a temperature dependence that is incompatible with simple spin-wave theory[61] (see **Box 4**). The full significance of the inelastic scattering data is more apparent in single-crystal measurements, discussed below.

**Figures 7a** and **7b** show INS data from single crystals of $\alpha$-RuCl$_3$ measured using time-of-flight methods[75,90]. Panel **a** shows the scattering associated with the zone-center (i.e. $\Gamma$ point) of the honeycomb lattice, both above and below $T_N$. In the ordered state the response shows two sharp peaks superposed on a broad continuum that continues to high energies. The sharp peaks arise from spin waves in the ordered state and disappear above $T_N$ leaving the continuum by itself. The continuum is temperature independent up to well over 100 K. The momentum dependence of the scattering is apparent in the upper panel of **Fig. 7b**, where the scattering at $T = 2$ K shows the spin waves as low-energy features with minima at the wavevectors $(\pm\frac{1}{2},0,0)$, and the continuum, centered at the $\Gamma$ point, is broad in both energy and momentum. It is apparent that the spin wave spectrum is gapped in the ordered state; presumably this is a signature of anisotropy in the interactions. The magnetic order in $\alpha$-RuCl$_3$ is suppressed by the application of an external magnetic field with magnitude $B_c \approx 7.5$ T along the in-plane <1 1 0> direction in reciprocal space (trigonal description), as shown in the phase diagram of **Fig. 5a**. In the resulting disordered state, the spin waves are again suppressed, while the continuum gains intensity at low energies. Although the energy gap at the Γ point softens as $B_c$ is approached, it appears that the continuum scattering is gapped above $B_c$; this is seen more easily in the line plots of scattering intensity vs. energy shown in Ref. 90. We note in passing that a gap in the disordered phase is presumably required to observe a quantized thermal Hall effect[21]. The magnetic interactions in α-RuCl$_3$ are very two dimensional, and the scattering plotted in **Fig. 7b** is integrated over a range of out-of-plane momentum values under the assumption that this enhances the counting statistics with no significant loss of information. However, it must be kept in mind that the ordered state itself is three dimensional, and by corollary so is the spin-wave spectrum.

The coexistence of sharp spin-waves and broad continuum scattering in the ordered state is reminiscent of the situation in systems of weakly coupled $S$ = 1/2 Heisenberg antiferromagnetic chains[119]. The natural excitations of the one-dimensional system are fractionalized spinons leading to a scattering continuum. At low temperatures the coupled chains order and the lowest energy excitations are three-dimensional spin waves; these coexist with the higher energy spinon spectrum. Above $T_N$ the spin-waves are gone, but the spinon scattering remains. This behavior is experimentally verifiable, for example in the quasi-one-dimensional magnet KCuF$_3$[120].

It is interesting to consider whether an analogous situation exists in α-RuCl$_3$. The observed continuum scattering at high energies can be compared to calculations of the response function for a pure[29] or perturbed[117] Kitaev model, and it is found that the overall extent and form of the scattering is consistent with the expectations for a Kitaev QSL. Moreover, the temperature dependence of the continuum scattering is also consistent with expectations for Majorana fermions in the Kitaev QSL[55,61,63,105]. Additional work is required to arrive at a definitive understanding of the INS measurements, since as of this writing there is still some disagreement concerning the correct Hamiltonian describing α-RuCl$_3$[116,122], the possibility that the scattering is better described by unstable magnons[123], and, given the similarity of response from the ferromagnetic and antiferromagnetic QSLs, whether there is ambiguity over the sign of the Kitaev term needed within a Kitaev description[90]. **Figure 8** shows some representative calculations of response functions for Hamiltonians containing both Kitaev and other terms. Despite the uncertainties, the preponderance of experimental evidence suggests that the continuum scattering seen in INS is a signature of fractional magnetic excitations, and these may be related to the excitations of a Kitaev QSL.

THz spectroscopy provides a high-resolution measurement of the response at zero momentum ($Q$ = 0), nicely complementing INS measurements, since due to kinematic constraints INS can access the Γ point of the quasi-two-dimensional Brillouin zone only for non-zero values of the out-of-plane momentum. THz spectroscopy has been used to obtain detailed measurements of the spin waves at $Q$ = 0[112,114,115] that provide crucial information about the full Hamiltonian describing α-RuCl$_3$, although it should be kept in mind that since the weak inter-planar interactions are apparently antiferromagnetic, in the ordered state the lowest gapped excitations at $Q$ = 0 may not represent the minimum spin gap associated with the 2D Γ point. Broadly speaking the results from THz spectroscopy are consistent with those from INS. **Figure 7c** shows measurements at $T$ = 2.4 K over a range of magnetic fields extending up to 15 T[113]. The spectra show sharp spin-wave peaks co-existing with a broad continuum at low fields, the vanishing of the spin-waves at $B_c$, and the emergence of a gapped mode at high fields with energy increasing linearly with $B$. A spin gap linear in $B$ is consistent with theoretical expectations for a system described by a Hamiltonian that is perturbed away from the Kitaev limit[102,117] (see **Box 3**). Such a mode has also been observed in high-field ESR experiments[109], albeit with an energy that appears somewhat different from that seen via THz spectroscopy.

The field dependence of the spin gap in $\alpha$-RuCl$_3$ as measured by NMR has been somewhat controversial. NMR is sensitive to the field induced suppression of magnetic order[88], but in the disordered state at least one group has reported gapless excitations[110], while another finds the gap growing as $B^3$ at high fields[111]. The latter result has been interpreted as arising from two-gauge-flux excitations such as are produced by a spin-flip in the pure Kitaev model, with the temperature dependence of the local susceptibility and spin relaxation providing evidence for the expected additional Majorana fermion.

For the pure Kitaev model the response function measured in INS contains contributions from pairs of static gauge-fluxes as well as one or more (odd numbers) mobile Majorana fermions[29]. In contrast, Raman scattering arises from process involving pairs of Majoranas, enabling an easier calculation of the intensity[30,124]. **Figure 7d** shows low temperature Raman scattering in $\alpha$-RuCl$_3$[107]. The continuum indicated in blue is evidently magnetic, cannot be explained by conventional two-magnon scattering, but has a strong resemblance to the expected scattering from a pure Kitaev model. Further analysis of the temperature dependence of the Raman signal provided additional evidence for the fractional nature of the underlying spin excitations[31].

**A half quantized thermal Hall effect in $\alpha$-RuCl$_3$**

A non-spectroscopic approach, such as thermal transport, is another promising way to detect the fractionalization of spin excitations. As discussed in the earlier section, the chiral edge state of the topological Kitaev QSL under magnetic field gives rise to a half-quantized thermal Hall effect $\kappa_{xy}/T$. The half-integer factor arises from fractionalization into Majorana fermions, as Majorana fermions carry only half of the degrees of freedom of canonical fermions. A thermal Hall effect close to a half quantized value $\kappa_{xy}/T = 1/2(\pi/6)(k_B^2/\hbar)$ was very recently reported for single crystals of $\alpha$-RuCl$_3$ with magnetic field close to the critical in-plane value, $B_c$ [91]. An unusually large thermal Hall effect $\kappa_{xy}/T$ with magnetic field perpendicular to the plane, $B_\perp$, (in-plane field $B_{//} = 0$) was discovered earlier in $\alpha$-RuCl$_3$ above $T_N$[125]. It was discussed as a possible signature of a half-quantized thermal Hall effect, masked by the long-range ordering below $T_N$. The in-plane field $B_{//}$ required to suppress the long-range ordering can be superposed on $B_\perp$ by tilting the magnetic field from the perpendicular direction. In the $B$-dependence of the tilted field data, a plateau-like behavior of $\kappa_{xy}/T$ as a function of applied field was observed as shown in **Fig. 5c**[91]. The $\kappa_{xy}/T$ value in the plateau region was independent of the tilting angle and close to the expected half-quantized thermal Hall effect $\kappa_{xy}/T = 1/2(\pi/6)(k_B^2/\hbar)$. The in-plane component of magnetic field $B_{//}$ in the plateau region was slightly larger than the critical field $B_c$, suggesting that the phenomena is closely linked to the $B$-induced suppression of the long-range magnetic ordering. At the time of writing, this striking observation[91] is too new to have been reproduced by other groups, and the full details have not been explored. The ultimate significance of the apparent fractional excitations in $\alpha$-RuCl$_3$ and the relationship of the field-induced disordered state to the Kitaev QSL are at the forefront of current research.

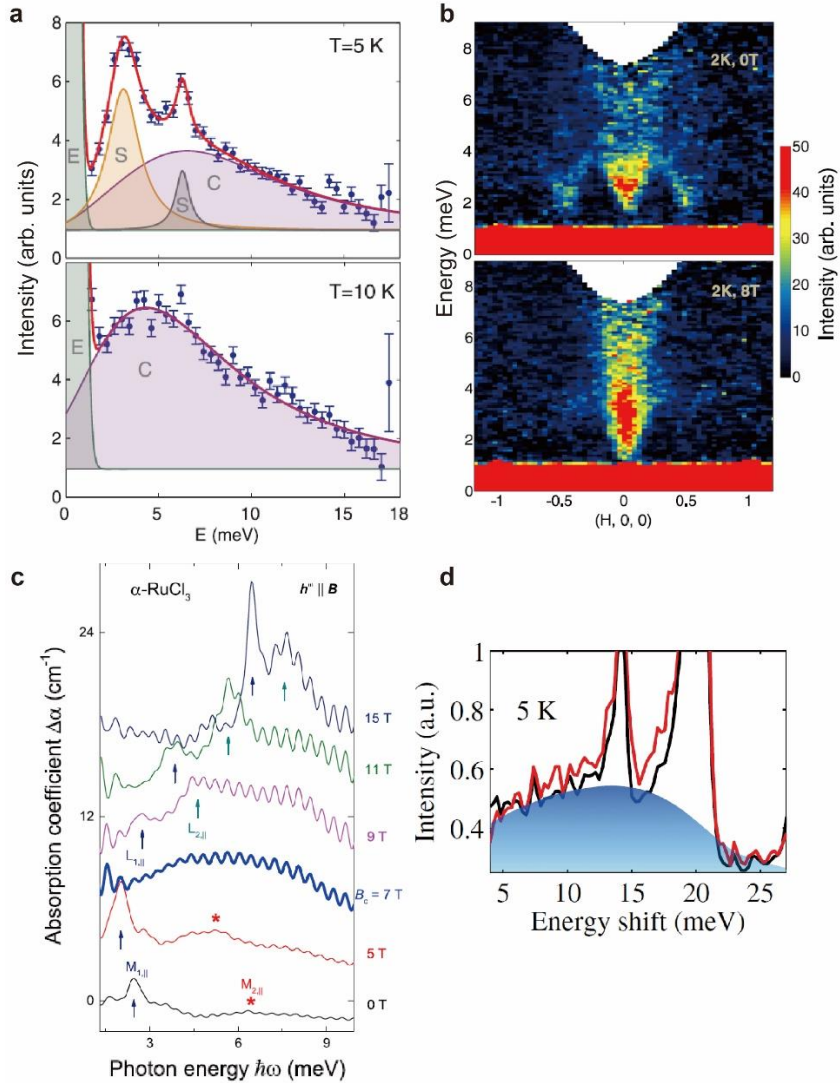

**Figure 7**. **Signature of fractional excitations in $\alpha$-RuCl$_3$. a,** Inelastic neutron scattering in single-crystal $\alpha$-RuCl$_3$ measured at temperatures of $T$ = 5 K (top) and 10 K (bottom)[75]. The data is integrated over a small reciprocal space volume centered at the $\Gamma$ point of the two-dimensional lattice. The letters designate the contributions from the elastic line "E", spin-waves "S", and continuum scattering "C". **b,** Inelastic neutron scattering is measured at $T$ = 2 K (top) in zero external magnetic field and (bottom) in a field of 8 T in the honeycomb plane, large enough to suppress the magnetic order[90]. The color bar denotes the relative intensity. **c,** THz spectroscopy measurements in $\alpha$-RuCl$_3$[113] in the presence of a magnetic field applied in the honeycomb plane, with the THz field parallel to the applied field direction. All measurements were carried out at $T$ = 2.4 K. The arrows indicate locations of excitations inferred from the data. **d,** Detail of Raman measurements in $\alpha$-RuCl$_3$ at $T$ = 5 K[107]. The blue shaded area represents the magnetic continuum scattering. (Panel **a** reproduced with permission from Ref. 75, panel **b** reproduced with permission from Ref. 90, panel **c** reproduced with mission from Ref. 113, and panel **d** reproduced with permission from Ref. 107).

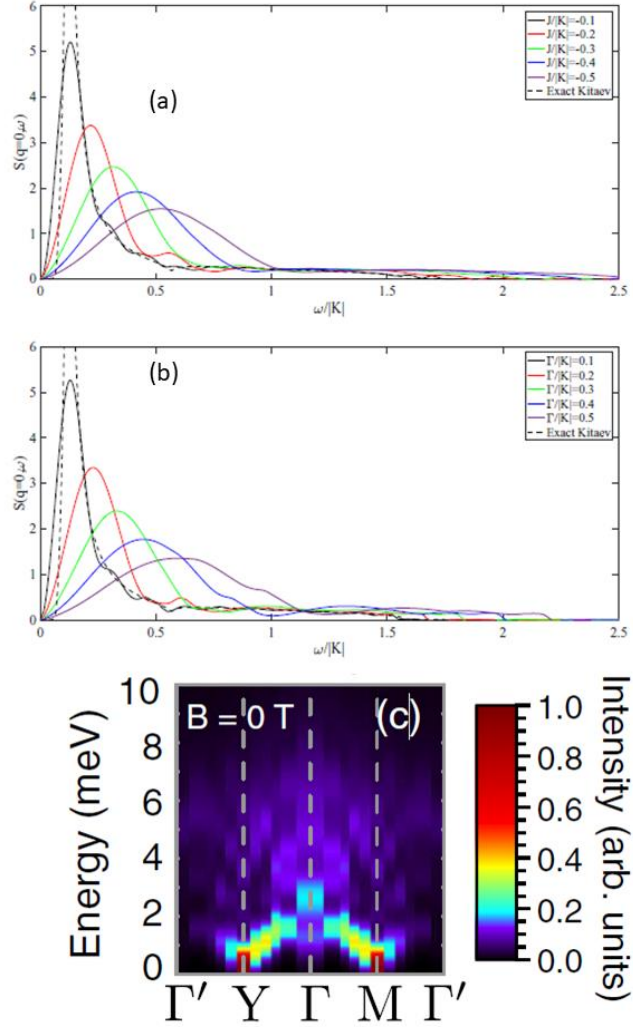

**Figure 8. Response functions in extended Kitaev models. a, b,** structure factor $S(Q=0, \omega)$ at the $\Gamma$ point calculated using perturbation theory from the pure Kitaev limit[121]. The Kitaev interaction assumed equal along all three directions is denoted by K. **a** shows the effect of a perturbing Heisenberg term of strength J, and **b** shows the effect of a perturbing off-diagonal exchange term $\Gamma$. The most obvious effects of either perturbation are a broadening and renormalization of the low energy peak in the spectrum that, in the pure Kitaev model, is a signature of the static fluxes **c,** Inelastic neutron scattering intensity calculated by exact diagonalization[117] for a model Hamiltonian including Kitaev, off-diagonal, and first and third neighbor Heisenberg terms. The parameter values used are representative of estimates proposed to describe $\alpha$-RuCl$_3$[123]. (Panel **a** and **b** reproduced with permission from Ref. 121, and panel **c** reproduced with permission from Ref. 117).

**Perspectives and summary**

The emergence of the exactly solvable Kitaev model a decade ago was a significant breakthrough in quantum magnetism. The impact of the Kitaev model was not confined to the theoretical community but soon found expression in correlated oxide physics, where the concept of the spin-orbital Mott insulator had just been established, and to topological physics, where the hunt for Majorana fermions became an important topic. This linkage grew into an interdisciplinary field whose main goal is the materialization of Kitaev quantum spin liquids. As seen in this review, the progress of the last few years is quite impressive. A number of honeycomb spin-orbital Mott insulators have been proposed to accommodate the essential ingredients of the Kitaev model. At the same time, a number of firm theoretical predictions for distinct physical properties of the Kitaev QSL, in particular those signifying the presence of Majorana fermions, have been proposed for the pure Kitaev model. Furthermore, a QSL (not yet identified as "Kitaev" QSL) ground state was established in $H_3LiIr_2O_6$ and the footprint of Majorana fermions may have been captured in $\alpha$-$RuCl_3$ under magnetic fields. Nevertheless, there is a gap between the materials and the theoretical models. The materials do not realize the pure Kitaev model but accommodate many other ingredients, such as additional exchange interactions, and these mask the manifestation of pure Kitaev physics. There is a need for realistic theories with additional ingredients to describe the QSL states and the elementary excitations. What kind of QSL are they? Is there any connection with the pure Kitaev model and other QSLs? Can the elementary excitations be described as Majoranas? There is also a need for more materials with the right ingredients for Kitaev physics. Can we have candidates without $Ir^{4+}$ and $Ru^{3+}$? The game has just started.

**Box 1. Spin-fractionalization in the Kitaev model**

The Hamiltonian of Kitaev model[21] reads as

$$H = - \sum_{<ij>_\gamma} K_\gamma S_i^\gamma S_j^\gamma$$

where $<ij>_\gamma$ stands for a $\gamma = x$, $y$ or $z$ type bond and the summation is taken over all honeycomb bonds. The coupling constants $K_\gamma$ carry a bond index and their value may differ on different types of bonds.

The model is characterized by infinitely many local conserved quantities, the integrals of motion, and is exactly soluble for any sign and relative strength of the couplings $K_\gamma$. The conserved quantities are flux operators $W_{1-6} = 2^6 S_1^z S_2^x S_3^y S_4^z S_5^x S_6^y$ defined individually around each hexagonal loop as a product of six spin operators $S_i^\gamma$ with $\gamma$ matching the index of the out-going bond. See **Fig. 1a**. The flux operators have quantized eigenvalues $W_{l-m} = \pm 1$, and commute with Hamiltonian and with each other. This allows each many-body eigenstate to be labelled by the conserved flux quanta through each hexagon and brings the Kitaev Hamiltonian to a block-diagonal form.

Alexei Kitaev's exact solution employs 'fractionalization' of the spin degrees of freedom via expressing $S = 1/2$ operators in terms of four different flavors of Majorana fermions[21]. The majoranas, first introduced by Ettore Majorana in high-energy physics, are neutral self-adjoint fermions being simultaneously particle and anti-particle. They can be constructed from the real or imaginary part of more common complex fermions. Hence, one complex fermion mode, described by $a$ and $a^\dagger$, give rise to two Majorana modes $c_1 = (a + a^\dagger)$ and $c_2 = i(a - a^\dagger)$. The spin 'fractionalization', expressed mathematically as $S_j^\gamma = \frac{i}{2} b_j^\gamma c_j$, together with the constraint $b_j^x b_j^y b_j^z c_j = 1$, preserves not only the $S = 1/2$ algebra but also the local two-dimensional Hilbert space. Most importantly, this choice of the Majorana representation transforms the Kitaev model into the fermionic form that explicitly and conveniently reflects the flux-operator ($W$) conservation law, which is the key to the exact solution. Namely,

$$H = -\frac{1}{4} \sum_{<ij>_{(\gamma)}} K_\gamma u_{ij}^\gamma c_i c_j$$

where the bond operators $u_{ij}^\gamma = b_i^\gamma b_j^\gamma$ with eigenvalues $\pm i$ commute with each other and with the Hamiltonian, and their product around a hexagon $W_{1-6} = u_{12}^y u_{23}^z u_{34}^x u_{45}^y u_{56}^z u_{67}^x$ determines the flux $W_{l-m} = \pm 1$. These commutation and conservation rules imply that $b_i^\gamma$ majoranas are constrained to the corresponding $\gamma$-type bond connected to site $i$ and are thus immobile. $u_{ij}^\gamma$ constitutes an emergent $Z_2$-gauge field and determines a phase of the nearest-neighbor tunneling integral of *c*-majoranas often termed as matter fermions. (See **Fig. 1d**.) In each flux sector, the gauge is fixed and the operators $u_{ij}^{\gamma}$ can be replaced by numbers $+i$ or $-i$. The ground state is flux free i.e. all $u_{ij}^{\gamma}$ are equal, as shown in **Fig. 1e,** and matter fermions can coherently propagate through the honeycomb lattice gaining the maximum kinetic energy. The corresponding dispersion is obtained by diagonalizing the quadratic (non-interacting) fermionic Hamiltonian after all link operators $u_{ij}^{\gamma}$ are replaced by $+i$. The obtained spectrum[21] is of Dirac-type, depicted in **Fig. 1f,** with all states appearing in pairs corresponding to positive and negative eigenvalues

$$E_{\boldsymbol{k}} = \pm\sqrt{\mathcal{E}_{\boldsymbol{k}}^2 + \Delta_{\boldsymbol{k}}^2}$$

where $\mathcal{E}_{\boldsymbol{k}} = 2[K_z - K_x \cos(\boldsymbol{k}\cdot\boldsymbol{a}) - K_y \cos(\boldsymbol{k}\cdot\boldsymbol{b})]$, $\Delta_{\boldsymbol{k}} = 2[K_x \sin(\boldsymbol{k}\cdot\boldsymbol{a}) + K_y \sin(\boldsymbol{k}\cdot\boldsymbol{b})]$, $\boldsymbol{k}$ and $\boldsymbol{a}(\boldsymbol{b})$ are quasi-momentum and honeycomb lattice vectors, respectively. The spectrum is gapless for weakly anisotropic coupling constants $K_\gamma$, and a gap opens when one of the couplings becomes larger than the sum of the remaining two $K_\alpha > |K_\beta + K_\gamma|$. In the gapless phase, the Dirac point can acquire a finite gap by time reversal symmetry breaking perturbations, e.g. external magnetic field induces a Majorana gap $\Delta_M \sim H^x H^y H^z / K^2$, where $H^\gamma$ ($\gamma = x, y$ or $z$) are the Cartesian components of the applied field and the exchange couplings are set to being equal ($K_x = K_y = K_z = K$) [21]. However, the power law of the induced gap vs. field is not universal, i.e. when the Kitaev model gets perturbed by additional couplings in actual compounds (Box 3), the field induced gap would then scale linearly with the applied field[102].

**Box 2.** Materials chemistry of Kitaev candidates

The crystal structure of Kitaev candidate materials consists of edge-sharing $IrO_6$ ($RuCl_6$) octahedra in a honeycomb-based network, which can be reconstructed from an ordered rock-salt (NaCl) structure. Consider first a NaCl-type $M''^{2+}O^{2-}$ (M": transition-metal atom), where all $M''^{2+}$ is octahedrally coordinated with $O^{2-}$. Each $M''O_6$ octahedron shares its edges with the neighboring twelve $MO_6$ octahedra. Viewing $M''^{2+}O^{2-}$ along the cubic (111) direction, we notice that the structure of $M''^{2+}O^{2-}$ consists of an alternating stack of the triangular $M''^{2+}$ planes and the triangular $O^{2-}$ planes. In this view, the $M''O_6$ octahedron consists of $M''^{2+}$ ions and the two $O^{2-}$ triangles directly above and below $M''^{2+}$ respectively. By replacing every pair of adjacent $M''^{2+}$ planes with an $A^+$ plane and $M'^{3+}$ plane, we have a layered $AMO_2$-type structure with triangular layers of $A^+$ and $M'^{3+}$, as can be seen in $LiCrO_2$ and $LiCoO_2$. The trivalent $M'^{3+}$ can be replaced by a 2:1 ratio of $M^{4+}$ and $A^+$. The 2:1 ratio of $M^{4+}$ and $A^+$ in the triangular plane can be realized when $M^{4+}$ forms a honeycomb lattice and $A^+$ occupies the center of $M^{4+}$ honeycomb. Thus formed $A_{1/3}M_{2/3}$ layers contain a honeycomb network of $MO_6$ octahedra connected by one of their six edges penetrating the M plane. The alternate stacking of an $A^+$-cation layer and an $A^+_{1/3}M^{4+}_{2/3}$ layer corresponds to the chemical formula $A_2MO_3$ (= $A(A_{1/3}M_{2/3})O_2$) as in $Na_2IrO_3$ and $\alpha$-$Li_2IrO_3$. The three-dimensional honeycomb structure of $\beta$-$Li_2IrO_3$ and $\gamma$-$Li_2IrO_3$ can be derived from the rock-salt structure, but the ordering pattern of $Li^+$ and $Ir^{4+}$ are different from the (111) ordering above. $\alpha$-$RuCl_3$ comprises of a similar honeycomb network of edge-sharing $RuCl_6$ octahedra but does not have any cations at the center of honeycomb plaquettes or between the honeycomb layers.

Single crystals of $Na_2IrO_3$ can be grown by a flux method[40]. The single crystal growth of $\alpha$-$Li_2IrO_3$ is not as easy as that of $Na_2IrO_3$, partly because of the existence of $\beta$- and $\gamma$-type polymorphs[52,53]. $\alpha$-$Li_2IrO_3$ single crystals can be grown by utilizing a vapor transport technique[65], while single crystals of $\alpha$-$RuCl_3$ can be obtained by a Bridgman technique[126], vacuum sublimation[50,69] or vapor transport technique[51]. The 2nd generation of Kitaev materials $A_3LiIr_2O_6$ (A' = Ag, Cu, H)[45-49] and $Cu_2IrO_3$[42] can be synthesized using a soft-chemical ion-exchange reaction, by soaking powder of $\alpha$-$Li_2IrO_3$ ($Na_2IrO_3$ for $Cu_2IrO_3$) in a molten salt or aqueous solution containing A' ions. Only polycrystalline powder is available for the 2nd generation materials at the time of writing. The single crystal growth of these materials is challenging but crucially important for further investigations.

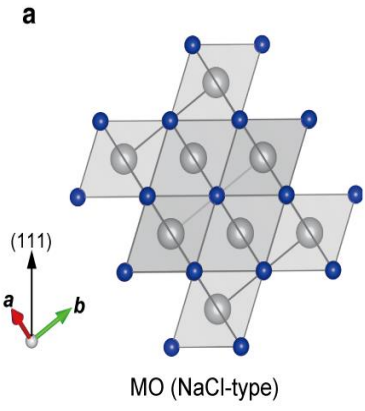
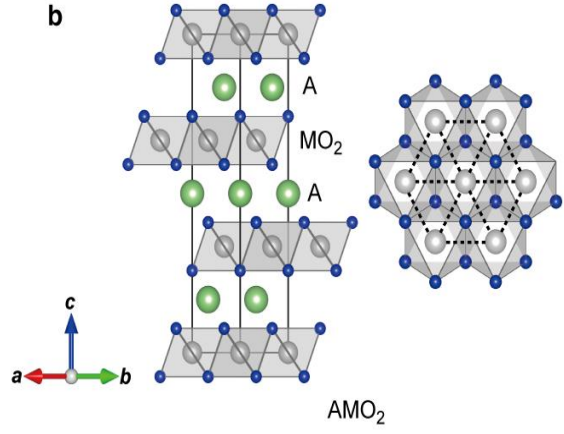
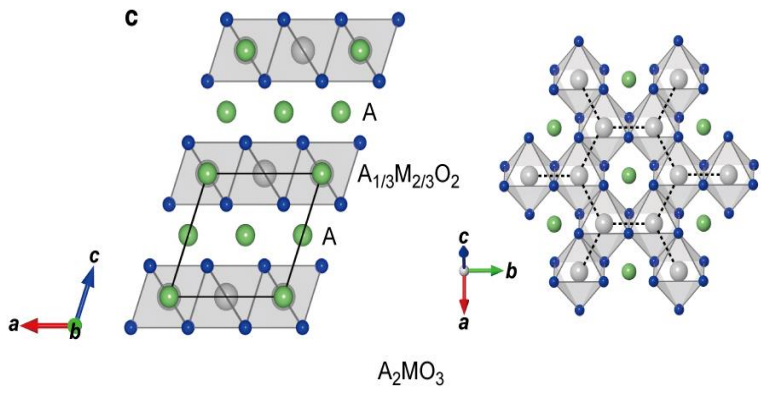

**Box 3 Non-Kitaev interactions and induced magnetically ordered phases**

The generic nearest-neighbor (NN) exchange Hamiltonian for the undistorted layered hexagonal $A_2IrO_3$ and $\alpha$-$RuCl_3$ compounds includes three symmetry allowed terms[23,76-80,127]

$$\mathcal{H} = \sum_{\langle ij \rangle_\gamma} \left\{ -K S_i^\gamma S_j^\gamma + \Gamma \left( S_i^\alpha S_j^\beta + S_i^\beta S_j^\alpha \right) + J \vec{S}_i \cdot \vec{S}_j \right\}$$

Here, $\langle ij \rangle_\gamma$ marks the three inequivalent NN bonds of the honeycomb lattice with $\gamma = x$, $y$, or $z$ and $S_i^{\alpha(\beta)}$ and $S_i^\gamma$ stand for the in-plane and out-of-plane Cartesian components of the $S = 1/2$ pseudo-spins in the Ir-$O_2$-Ir plane, labelling the $J_{eff} = 1/2$ Kramers pairs.

The first Kitaev term ($K$) originates from the combined effects of the anion mediated super-exchange and Hund's coupling, as discussed in the main text. All available theoretical methods, including perturbation theory, *ab-initio* and quantum chemistry calculations, suggest that the Kitaev term dominates the other two terms and is ferromagnetic ($K > 0$)[23,57,78]. The off-diagonal symmetric anisotropy, the $\Gamma$ term, is of next-to-leading order and requires both direct *d-d* and anion mediated *d-p* electron transfer[77]. The isotropic Heisenberg term is of antiferromagnetic type ($J > 0$). It has the smallest strength since it predominantly originates from the weak direct *d-d* hybridization.

The above minimal model has been extensively studied by analytical and numerical techniques, and the extremely rich phase behavior is now understood[76,77,116,128,129]. For $\Gamma = 0$, the model supports four magnetically ordered phases with collinear spin patterns of ferro, antiferro, stripy, and zigzag types[76,116]. Most remarkably, the phase diagram includes a finite stability window for the QSL phase close to the Kitaev limit ($J = 0$). Finite $\Gamma$ further enriches the phase diagram by adding non-collinear and incommensurate spiral phases[77].

The experimentally detected zigzag magnetic ordering in $Na_2IrO_3$ and $\alpha$-$RuCl_3$ appears on the theoretical phase diagram of the minimal nearest-neighbor model for antiferromagnetic Kitaev coupling ($K < 0$). However, it is widely believed that the Kitaev term in these compounds is of ferromagnetic type instead. It has therefore been proposed that exchange couplings beyond nearest-neighbor are responsible for stabilizing the zigzag order[55,81] observed in $Na_2IrO_3$ and $\alpha$-$RuCl_3$. Indeed, the *ab-initio* parametrization of further-neighbor interactions indicate that they become sizable in these compounds[57].

The non-cubic crystal field arising from the distorted octahedra mixes $J_{eff} = 1/2$ and $J_{eff} = 3/2$ states. The ground state remains doubly degenerate, protected by time-reversal symmetry, and can be still described by pseudo-spin $J_{eff} = 1/2$. However, as the associated wave function becomes modified, the destructive quantum interference, mentioned above, is no longer exact and the isotropic Heisenberg term becomes larger. Distortion-induced lowering of the crystal symmetry also gives rise to other non-Kitaev interactions, such as antisymmetric anisotropy known as the Dzyaloshinsky-Moriya term.

**Box 4. Inelastic neutron scattering measurements in polycrystalline materials**

The cross-section for INS measurements in single crystals of collinearly ordered magnetic systems consists of a sharp peak at the position of allowed spin-wave modes. The dispersion (i.e. wavevector dependence of the energies) and strength of the peaks can be analyzed to infer the parameters relevant to a proposed model Hamiltonian. Panel **a** shows a representative spin wave model for a single crystal zigzag-ordered honeycomb lattice, using the Hamiltonian parameters proposed for $\alpha$-RuCl$_3$[116,117]. The relative intensity expected in a single-crystal INS experiment is represented by the color as defined in the color bar (right). The color bar convention is used in all five panels.

When measurements are carried out on powders the orientational information is lost and the scattering is proportional to a weighted density of states. This yields less information but can still be incredibly useful. Panel **b** shows neutron scattering intensity measured in a powder of Na$_2$IrO$_3$[55], with scattering at energies below 2 meV suppressed for clarity. As discussed in the text it is experimentally very challenging to get high quality INS data in the iridate materials. Compounds based on Ru do not suffer from this complication so measurements with much better statistics are possible, as illustrated in panel **c** showing INS data from a powder of $\alpha$-RuCl$_3$[61]. Note that there is no normalization between the data sets plotted in panels **b** and **c**.

The powder scattering in both Na$_2$IrO$_3$ (panel **b**) and $\alpha$-RuCl$_3$ (panel **c**) shows a low energy feature with a threshold shape that is concave towards the origin. This was interpreted as a signature of underlying zigzag order since, for most plausible Hamiltonians describing the system, spin waves from the stripy ground state yield a convex threshold shape[55,61]. Representative calculations of the powder averaged scattering model (Ref. 61, supplemental materials) is shown for both the zigzag (panel **d**) and stripy (panel **e**) models. The threshold is shown by the white arrow in each case. The measured shape of the scattering threshold indicates zigzag order in both Na$_2$IrO$_3$ and clearly the low-momentum scattering threshold is much more crisply defined in $\alpha$-RuCl$_3$.

The complete scattering in $\alpha$-RuCl$_3$ contains an additional feature at higher energies[61] that shows up near 6 meV in panel **c**. When the temperature is increased to $T > T_{mag}$ (near 14 K in polycrystalline material) the lower energy spin-wave scattering loses definition and softens dramatically. One would expect the scattering from a higher energy spin wave mode to be greatly diminished at temperatures larger than $2T_{mag}$. This is not the case for the feature seen in $\alpha$-RuCl$_3$ whose intensity persists to temperatures of up to 100 K or more. This scattering arises from the powder average of the continuum scattering seen near the $\Gamma$ point (see main text, and supplementary material of Ref. 75 for a more detailed discussion). The overall energy width and $T$-dependence of this feature was seen to be consistent with scattering expected from fractional excitations[61].

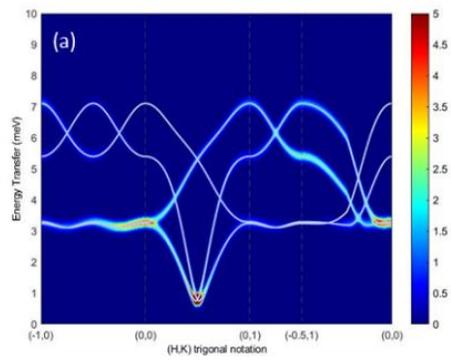
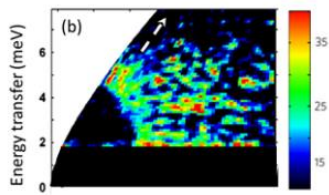
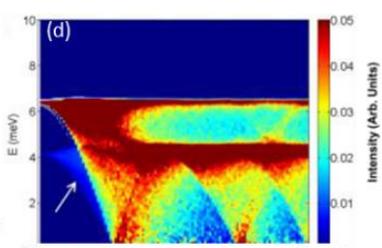
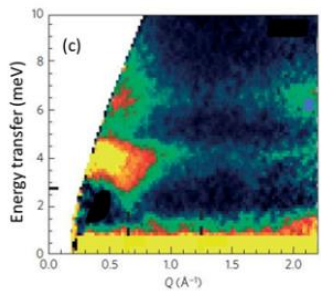
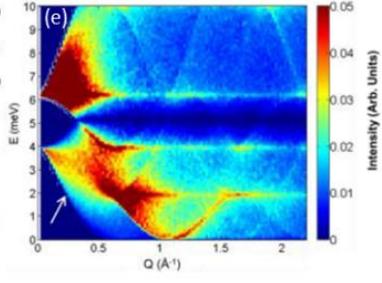

*Acknowledgements*

We would like to thank A. Smerald for his critical reading of this manuscript. HT, TT and GJ acknowledge support from Alexander von Humboldt foundation. We thank A. Banerjee for preparing the figure shown in panel (a) Box 4. HT was supported by Japan Society for the Promotion of Science (JSPS) KAKENHI (17H01140, JP15H05852, JP15K21717). SN was supported by the US Department of Energy, Basic Energy Sciences, Scientific User Facilities Division.

*Author contributions*

All authors contributed the preparation of this manuscript.

*Competing interests*

The authors declare no competing interests.